# Emergent topological magnetism in Hund's excitonic insulator

R. Okuma[1,2*], K. Yamagami[1,3*], Y. Fujisawa[1], C. H. Hsu[1], Y. Obata[1,4], N. Tomoda[1], M. Dronova[1], K. Kuroda[2], H. Ishikawa[2], K. Kawaguchi[2], K. Aido[2], K. Kindo[2], Yang-Hao Chan[5], H. Lin[5], Y. Ihara[6], T. Kondo[2], Y. Okada[1]

*equal contribution

**E-mail:** rokuma@issp.u-tokyo.ac.jp, yoshinori.okada@oist.jp

[1]*Okinawa Institute of Science and Technology Graduate University, Onna-son, Okinawa, 904-0495, Japan*
[2]*Institute for Solid State Physics, University of Tokyo, Kashiwa, Chiba 277-8581, Japan*
[3]*Japan Synchrotron Radiation Research Institute, Hyogo, 679–5198, Japan*
[4]*Department of Physics, Kanazawa University, Ishikawa 920-1192, Japan*
[5]*Institute of Physics, Academia Sinica, Taipei 115201, Taiwan*
[6]*Department of Physics, Faculty of Science, Hokkaido University, Sapporo 060-0810, Japan*

Analogous to the charged electron-electron pair condensation in superconductors, an excitonic insulator (EI) represents Fermi surface instability due to spontaneous formation and condensation of charge-neutral electron-hole pair (exciton). Unlike in superconductors, however, the charge-neutral nature of exciton makes probing emergent EI phase via macroscopic physical properties generally difficult. Here, we propose a van der Waals coupled antiferromagnetic semiconductor GdGaI (GGI) as a new material category leading to emergent multi-q magnet intertwined with spontaneous exciton formation/condensation. Before excitonic band hybridization, a simple picture for the parent electronic state consists of electron (Gd-derived 5$d$) and hole (Ga-derived 4$p$) delocalized bands, together with Gd-derived 4$f$ localized antiferromagnets with $S = 7/2$ classical nature. Through intra Gd atom 4$f$ – 5$d$ Hund's coupling, a notable finding is the emergent minimum length scale (~2a) Skyrmion-like spin texture resulting from spontaneous condensation/formation of spin-polarized exciton with BCS-BEC crossover phenomenology. This discovered platform is promising for realizing valuable quantum matter on the nanoscale; our finding will provide significant insight into designing the atomic scale topological magnetism out of itinerant systems.

## Introduction

**Interest in excitonic insulator (EI)**

Superconductivity is a condensation of Cooper pairs emerging from effective attractive electron interactions. In semiconductors or semimetals with an insufficient screening of Coulomb interaction, a bound state between electron and hole can spontaneously condense into an insulating phase called excitonic insulator (EI)[1]. The beautiful analogy of EI to superconductivity immediately encounters the problem of inevitable phonon condensation with the excitonic condensate in crystalline materials, which makes it difficult to clearly distinguish the EI from a more common charge density wave (CDW) insulator[2]. Regarding this controversy, semiconductors are a better platform to preclude the instability of fermi surfaces such as nesting[3]. Furthermore, while recent efforts demonstrate the importance of realizing EI states by separating electrons and holes in space by engineering heterostructures[3-6], the charge-neutral and spinless nature of the existing EI candidates limits experimental probes for the nontrivial features of EI so far[5-7].

**Exotic EI search by implementing antiferromagnetism**

Introducing magnetism in the EI may result in detectable orders qualitatively different from the CDW order, as the magnetic interaction is arguably the key to unconventional superconductivity in cuprates and iron-based superconductors[8]. Analogous to the unconventional superconductivity[9], EI has two regimes where the electron-hole pair formed locally as in the Bose-Einstein condensation (BEC) or in the momentum space Bardeen–Cooper–Schrieffer (BCS) mechanism, which corresponds to semiconductor and semimetal, respectively (**Fig. 1a**). Exciton formation dressed in spin degrees of freedom will engender a new condensed matter platform to search for exotic physics of magnetically coupled exciton condensation with tunable BCS-BEC phenomenology. For example, in the BCS regime, exciton formation and condensation happen simultaneously in association with long-range magnetic orders such as ferromagnetism[10]. In contrast, broad temperature regions hosting exciton formation without coherence in the BEC regime may lead to a spin liquid-like state. The BCS-BEC crossover is a fertile playground where exciton formation and condensation occur successively. This situation is roughly analogous to the underdoped high-$T_C$ cuprates, where multiple characteristic temperatures exist, including the coherent superconducting temperature $T_C$ and the temperature at which electrons are bound without coherence between pairs[11]. Intertwining magnetism into BCS-BEC crossover phenomenology will open a rich quantum material platform. However, such a platform has been missing so far. A two-dimensional narrow gap semiconductor hosting frustrated magnetism is a fascinating target to develop as a parent electronic state leading to exotic magnetic EI.

**Novelty and significance of GdGaI**

Here, we report on GdGaI (GGI) as a promising candidate for emergent excitonic topological antiferromagnetism within the BCS-BEC crossover region. The GGI is isostructural to the recently discovered exfoliative heavy fermion system CeSiI[12-14], whereas a larger magnetic moment of $S = 7/2$ from Gd and reduced number of valence electrons by replacing Si with Ga results in distinctive, exciting findings in this study. The structure derives from the EI candidate 1$T$-TiSe$_2$ (**Figs. 1b-c**) by doubling the unit cell along the $c$-axis[15,16]. In contrast to the inversion symmetric TiSe$_6$ octahedra in TiSe$_2$, each Gd$^{3+}$ ($4f^75d^0$) is surrounded by three Ga$^{2-}$ ($4s^24p^3$) and three I$^-$ ($5s^25p^6$) to form a polar molecular geometry in GGI. (**Fig. 1c**). The density functional theory (DFT) calculation predicts similar parent electronic states between TiSe$_2$ and GGI, where electron ($d$-orbital) and hole ($p$-orbital) bands dominate near the Fermi level at Γ and M points, respectively (**Fig. 1d** and **Supplementary note 1**). The delocalized band with excitonic instability may couple to the sizable localized magnetic moment of the Gd 4$f$ orbital on a frustrated triangular lattice of GGI. Notably, this study relies on our first successful growth of large-scale, high-quality single crystals of GGI (**Figs. 1e-f**). See also **Extended Data Table 1** and **Method section** for further details about crystal structure and crystal growth. By a comprehensive study of angle-resolved photoemission spectroscopy (ARPES), magnetometry, X-ray magnetic circular dichroism (XMCD), and nuclear magnetic resonance (NMR) (**Figs.1g-h**), GdGaI is demonstrated as a promising EI hosting a Skyrmion-like noncoplanar triple-$q$ magnetic order (topological magnet) through ferromagnetic Hund's rule coupling between 4$f$ and 5$d$ orbitals of Gd atoms.

## Results and Discussion

**Excitonic band deformation: comparison between 300 and 14 K**
From ARPES measurements, semiconducting band nature at 300 K and underlying band folding with $q = a^*/2$ at 14 K are presented first. The photon energy (He-Iα, $h\nu = 20.218$ eV) is initially selected to simultaneously capture the delocalized electronic nature around momentum Γ and M. While finite $k_z$ dispersion is not excluded, the effect is ignored within the main scope of this study (**Supplementary note 2**). By comparing the band dispersion between 300 K and 14 K, the band top around the Γ point is deformed within a few hundred meV from $E_F$ (**Fig. 2a**). For further comparison, the energy distribution curve (EDC) at the Γ point is compared between 300 K and 14 K (**Fig. 2b**). Suppressed spectral weight at the high binding energy of $E - E_F \sim -2.6$ eV at 14 K compared to 300 K (see the arrow) implies possible deformation of Gd-derived 4$f$ electrons, of which detailed investigation is left for future study. The electronic state outside these energy windows is basically intact. Therefore, the band deformation near $E_F$ is intrinsic rather than an irrelevant charging effect. As expected, the band deformation near Γ is associated with an alternation of band curvature from being electron-like (at 300 K) to being hole-like (14 K) near the M point (see **Figs. 2c and d**). By symmetrizing EDC against $E_F$, which is a conventional way to eliminate the thermal broadening effect[17], supports the presence of a semiconducting band structure with a gap of ~ 80 meV between the top of the hole band and bottom of the electron band at 300 K (see **Fig. 2e**). In contrast, the top of h-bands at the Γ and M points is at ~200 meV below $E_F$ at 14 K (**Fig. 2f**). This identical band top position supports that the hole band at the M point is a replica of the hole band at Γ point, in association with a propagation vector $q = a^*/2$. The weaker band signature of the hole band at the M point is also consistent with its replica origin. These band deformations from 300 K to 14 K are consistent with the schematic shown in **Fig. 1g**. Before introducing further data, an important fact to be stressed here is the large ~ 80 meV scale to be overcome to meet electron and hole bands, which is a stark contrast to the semimetal case TiSe$_2$. This ~ 80 meV energy scale is too large to be interpreted as a conventional-CDW-based spontaneous excitonic band hybridization. Thus, the large attractive Coulomb energy scale $U$ is naturally introduced to interpret the phenomenology we show hereafter.

**Two characteristic temperatures and an emergent camelback-shaped band**
The more detailed temperature evolution of the band structure is shown next, focusing on the Γ and M points consistently (**Figs. 2g-k**). By cooling sample, the band top of the Γ point shifts to a lower energy (**Fig. 2g**) simultaneously with the change in the band curvature at the M point (**Fig. 2h**). This trend is captured more systematically by visualizing the temperature evolution of symmetrized EDC spectra at the Γ and M points (**Figs. 2i and j**). As seen in **Figs. 2i-j**, the gap evolution of hole bands (main at Γ and replica at M) is consistently observed with the same energy scale (see dotted lines in **Figs. 2i-j**). From the temperature evolution of band dispersion near the Γ point (**Figs. 2i and k**), two characteristic temperatures, ~ 200 K and ~ 40 K, are recognized. Above 200 K, the hole band top position does not shift with cooling, whereas a gradual gap increase is recognized below 200 K. We stress that the temperature region below ~ 40 K is exceptional; a more pronounced replica hole band feature is recognized at both M and Γ points (see **Figs. 2i-k**). Furthermore, by plotting temperature-dependent dispersion extracted from peak fitting of EDC, the minimum band gap at finite momenta ± $k_H$ (see the shaded area in **Fig. 2k**) becomes evident by forming a camelback-shaped band structure centered at the Γ point. These observations imply that electron and hole bands try to hybridize spontaneously, gradually below 200 K, by compensating for more than 80 meV scale energy whereas the electron-like curvature is not apparent on the hole band. Strikingly below ~ 40 K, the electron-like signature on the hole band clearly appears with the emergent camelback band.

**Laser ARPES focusing on camel back shaped band**
Since the band around the M point is a replica in nature, a deeper understanding of band structure focusing predominantly around the Γ point is reasonable (**Fig. 1g**). A laser ARPES measurement is one of the best ways to investigate the band structure near the Γ point with high momentum and energy resolutions. The capability to probe momentum-dependent orbital character is also an advantage of the polarization-dependent study. In this study, the photon energy ($h\nu = 6.994$ eV) with selecting $s$- and $p$- polarizations are employed (**Fig. 3a**). At 7 K, the $p$-polarized light result reveals the presence of dramatic intensity change across characteristic momentum $|k_H|$ (**Fig. 3b**). In stark contrast, with the use of *s*-polarized light, stronger intensity is seen for the small momentum region $|k| < k_H$ (**Fig. 3c**). Notable fact is that this momentum dependent intensity between *s*- and *p*- polarized light is anticorrelated (see **Figs. 3b and c**), which is consistent with hybridization between *d* (electron) and *p* (hole) orbitals as schematically represented in **Figs. 1d and g**. The temperature dependence of ARPES with *p*-polarized light (**Fig. 3d**) visualizes the systematic gap evolution across momentum ~$k_H$ (see **Supplementary notes 1 and 2** for further details of selection rules).

**The connection between the band deformation and the evolution of ferromagnetic interaction**
The most crucial information for drawing the essential picture in this study is provided from a quantitative comparison of the band deformation and magnetic temperature evolution. Based on the ARPES data presented up to this point, focusing at Γ point, is simply drawn in **Fig. 4a**. Notably, the temperature evolution of antiferromagnetism on triangular-based lattice also represents vibrant behavior involving multiple characteristic temperatures (**Fig. 4b**). By introducing the energy scale $|E_\Gamma(T)|$ and $|E_\Gamma(T) - E_{kH}(T)|$, the degree of gradual (yellow area) and abrupt geometrical band deformations (blue area) are compared with reduced susceptibility $(\chi-\chi_0)T$ and magnetization $M$, respectively. Here, $\chi_0$ is the background signal from the temperature region where Curie Weiss (CW) law captures nicely (see the black line in **Fig. 4b**). In **Fig. 4c**, the onset of gradual band gap evolution evidenced by $|E_\Gamma(T)|$ across $T^*$ (upper panel) corresponds to the anomaly seen on $(\chi-\chi_0)T$ (lower panel). If a CDW played a dominant role (short- or long-range) below $T^*$, a downward deviation of $(\chi-\chi_0)T$ from the CW fit would be observed with the cooling sample. However, this contradicts the experimental upward deviation (**Fig. 4b** and **Fig. 4c** lower panel). Instead of the CDW origin, the natural interpretation is the magnetic anomaly below $T^*$. Regarding the length scale of the magnetism, the convex downward shape of $|E_\Gamma(T)|$ below $T^*$ points to the evolution of the short-range order as seen in nematic fluctuation in iron superconductors[18], rather than the onset of static long-range order. Since reduced susceptibility $(\chi-\chi_0)T$ measures the ferromagnetic spin-spin correlation[19], an enhanced signal right below $T^*$ indicates either enhanced ferromagnetic fluctuation or increased magnetic moment out of the parent antiferromagnetism. With cooling sample further, weak-field magnetometry 28 K shows the evolution of net moment below critical temperature $T_c$ (upper panel), in accord with the evolution of $|E_\Gamma(T) - E_{kH}(T)|$ (lower panel). The natural expectation that arises here is the existence of an intriguing mechanism connecting the near $E_F$ band deformation and established net moment.

**The Hund's rule coupling**
We explain the significant role of Hund's rule coupling in interpreting our observation. The 4*f* orbitals of Gd are intensely localized, unlike in heavy fermion compounds. Through so-called Hund's rule coupling, the energy of the 5*d* electron band with spin parallel to the 4*f* moment is lowered when ferromagnetic coupling is considered. To verify this picture, the X-ray Magnetic Circular Dichroism (XMCD) experiment is one of the most straightforward ways due to its capability to investigate orbital-dependent magnetism (see **Method section** for further details). From XMCD at the $M_{2,3}$ edge, which is sensitive to the magnetic state from 5*d* orbital, the energy split between spin up and down bands near $E_F$ at 15 K is clarified (**Fig. 4e**). Based on the XMCD data, energy scale of spin splitting of 5*d* bands near $E_F$ is evaluated as $\Delta E = 270$ meV (see **Fig. 4e**). This observation is due to the internal field from the Hund's rule coupling between the conduction 5*d* band and the localized 4*f* moment rather than the Zeeman splitting from the applied field. In addition to the $M_{2,3}$ edge measurement, the $M_{4,5}$ based measurement confirms the picture of ferromagnetically coupled 4*f* and 5*d* orbitals within the Gd site (**Extended Data Figs. 1-2**). Notably, compared to 4f orbital, minor net moment but larger energy scale band splitting is induced in the delocalized 5*d* orbital due to Hund's rule spin splitting. Therefore, the overlap between the Ga-4*p* and Gd-5*d* bands is controlled.

**The BCS-BEC cross-over in the itinerant channel**
Through Hund's coupling, the striking analogy emerges between incoherent/coherent exciton (delocalized channel) and incoherent/coherent magnetism (localized channel). For $T_c < T < T^*$, an exciton pair density is assumed to increase gradually with cooling, while their coherence is not obtained. This is analogous to the BEC picture, better captured by pairing in r-space without their overlapped wave function (**Fig. 4f** middle). Interestingly, corresponding to the incoherence of the exciton pair, the ferromagnetic spin correlation is also short-range. On the other hand, regardless of the expected subtle increase in exciton pair density for $T < T_c$, the fingerprint of coherent electron-hole hybridization in k-space becomes evident by abruptly emergence of camelback band (see **Fig. 4a**). This is analogous to a BCS-like picture, in the sense that pair formation/condensation is better characterized in k-space (**Fig. 4f** right). Strikingly, corresponding to the coherence of the exciton pair, magnetism is long-range weak ferromagnetism.

**Emergent Skyrmion-like lattice with minimum (~2$a$) length scale**

Finally, the emergent Skyrmion-like magnetic order with a minimum length scale ~2a is shown in the localized channel. Conventional Heisenberg-type interactions on a uniform triangular lattice generically stabilize single-$q$ spiral states under zero-field, even with further neighbor interactions[20-22]. However, if the primary energy gain is through excitonic band deformation, localized magnetism and delocalized excitonic channels should cooperatively represent triple-q orders. Since this is through ferromagnetically coupled intra-atomic Hund's rule coupling, ordering for localized magnetic channels can be easily locked by ~2$a$ characteristic length scale identical to one in excitonic delocalized channels. As in **Fig. 5a**, when the band top energy of unfolded hole constant energy contour is considered (blue), the degree of hybridization with triple-q folded ellipse-shaped electron constant energy contour (red) becomes minimum along ΓM direction (shaded area around Γ). Consequent momentum-dependent hole band top, based on laser ARPES data at 7 K, is consistently visualized (**Fig. 5b**). Furthermore, the emergence of a triple-$q$ magnetic state is convinced by nuclear magnetic resonance (NMR) experiments, which have the advantage of distinguishing single-$q$ and multiple-$q$ magnetic structures as a local probe (experimental details and analysis are described in **Methods section** and **Extended Data Tables 2-3** and **Extended Data Figs. 3-5**). The NMR spectrum uniquely nails down the ground state magnetic structure, as shown in **Fig. 5c-d**. While the unit cell contains two Gd- triangular sheets in the unit cell, they turned out to be in-phase in the c-direction, suggesting ferromagnetic interaction at the nearest Gd-Gd vertical bond. Therefore, consideration of the single triangular layer should capture the essence. The refined magnetic structures host a noncoplanar triple-$q$ structure with ~2$a$ periodicity, which can explain the presence of weak ferromagnetism due to spin canting below $T_c$ (**Fig. 4b**). Interestingly, focusing on the highlighted area in **Fig. 5c-d**, the noncoplanar spin structures can be mapped to the all-out (**Fig. 5e**) and all-in (**Fig. 5f**) spin structure on a tetrahedron, respectively. Depending on out-of-plane magnetic fields, one of the two stable structures with opposite net moment directions is realized while these two structures (**Fig. 5c** or **Fig. 5d**) can coexist at zero magnetic fields. It should be stressed that these effective *all-out* and *all-in* localized magnetic structure has non-zero scalar spin chirality <$S_i \cdot S_j \times S_k$> in every triangular plaquette, allowing the *all-out* structure to be topologically distinguished from the *all-in* structure due to their opposite sign of chirality. In GGI, the emergence of an exotic magnetic state is also speculated for $T_c < T < T^*$. With negligible magnetic anisotropy assumed in this temperature region, spins may uniformly rotate by keeping the local tetrahedral spin configuration as expected for the local strong correlation in the BEC excitonic state, which maintains the sign of chirality. This topological nature anticipates the emergence of a chiral spin liquid state in $T_c < T < T^*$ without dipole long-range order[23,24].

## Summary and Impact

In summary, we take a novel approach to realize the magnetic manifestation of spontaneous exciton formation and condensation in the van der Waals coupled semiconductor GdGaI. We discovered the new phenomenology intertwining emergent weak ferromagnetism and exciton formation/condensation in the BCE-BEC crossover region through Hund's rule-driven band deformation. The striking consequence is the realization of an emergent ~2$a$ scale minimum skyrmion-like lattice associated with exciton condensation. Shedding light on exciton condensation is a new material design principle to obtain exotic multiple-$q$ magnetic structures at zero-field, which has been considered as a highly nontrivial task in inversion symmetric magnetic systems over the years. Towards realizing useful quantum matter in the nanoscale, our finding will provide a novel insight into designing the atomic scale topological magnetism in itinerant systems. Our finding also provides a wealthy platform for searching a variety of intriguing phenomena, including the search for tunable emergence of topological magnetism intertwined with magnetic exciton formation/condensation and their optical/electrical/mechanical control in ultrathin film/flake-based devices.

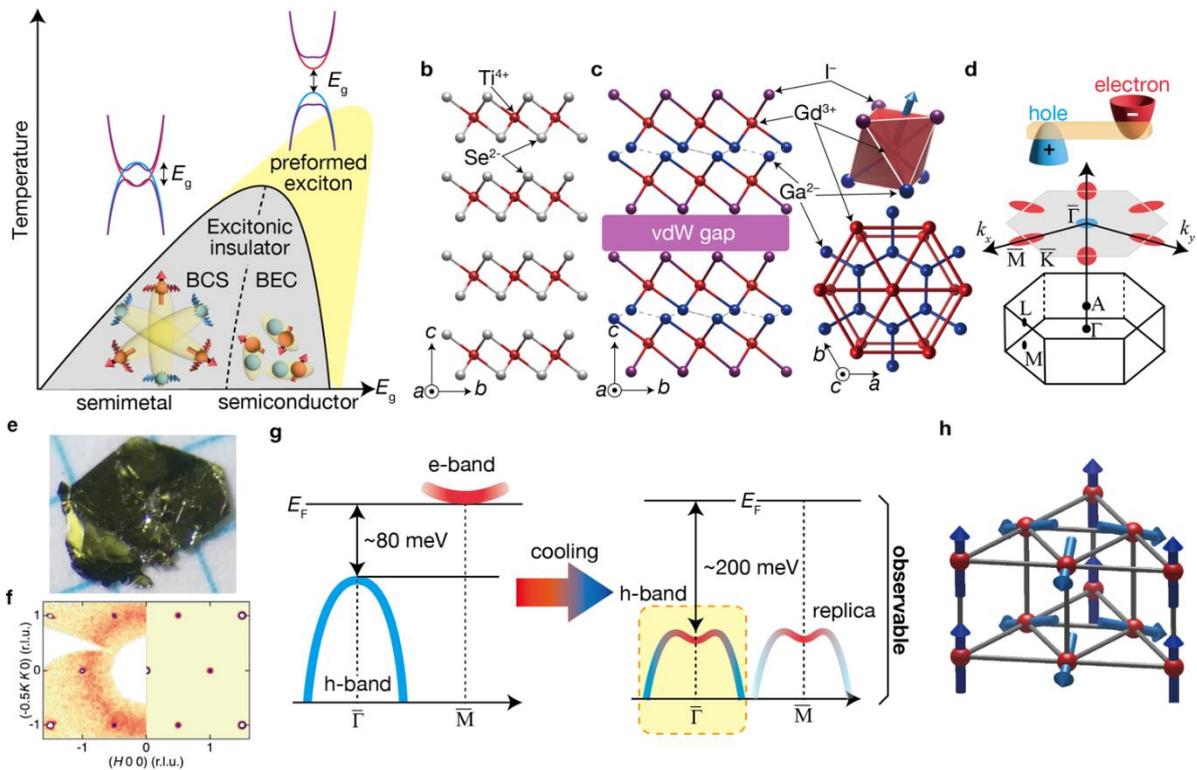

**Fig.1 | The van-der Waals coupled GdGaI (GGI) as a novel platform hosting topological magnet and excitonic insulator (EI) state.**

**a,** The concept of EI with implementing the spin degree of freedom. The magnetic moment of itinerant electrons is shown in a red arrow. The direction of the spin is fixed concerning the electron's momentum, and the excitonic wavefunction's coherence directly affects the spin fluctuation. **b,** The crystal structure for spinless EI candidate TiSe$_2$. For TiSe$_2$, red and grey spheres represent Ti and Se atoms, respectively. **c,** The crystal structure of GGI is the main focus of this study. Red, blue, and purple spheres represent Gd, Ga, and I, respectively. Gd has a magnetic moment of $S = 7/2$. **d,** The schematic image of the band structure (top), fermi surface (middle), Brillouin zone (bottom) of 1T-TiSe$_2$, and GGI. See the main body for details. **e,** An optical microscope image of a single crystal GGI. **f,** The observed (left) and calculated (right) the X-ray diffraction (XRD) pattern of the *HK*0 plane of GGI. **g,** The overall picture of the change in the band structure for GGI with the cooling sample. At room temperature, an indirect band gap of 80 meV is opened between the Ga-4*p*-derived hole band at Γ point and the Gd-5*d*-derived electron band at M point, with the Fermi level touching the bottom of the electron band (left). See **Supplementary Note 1** for further detailed band calculation. At the lowest temperature, hybridization between the two bands shifts the maxima of the hole band to lower energy by 120 meV and deforms it into a camelback-shaped band. In contrast, at the M point, the electron band is fully gapped out to higher energy and replaced with a replica band of the hole band (right). **h,** The noncoplanar triple-*q* magnetic order realized in GGI.

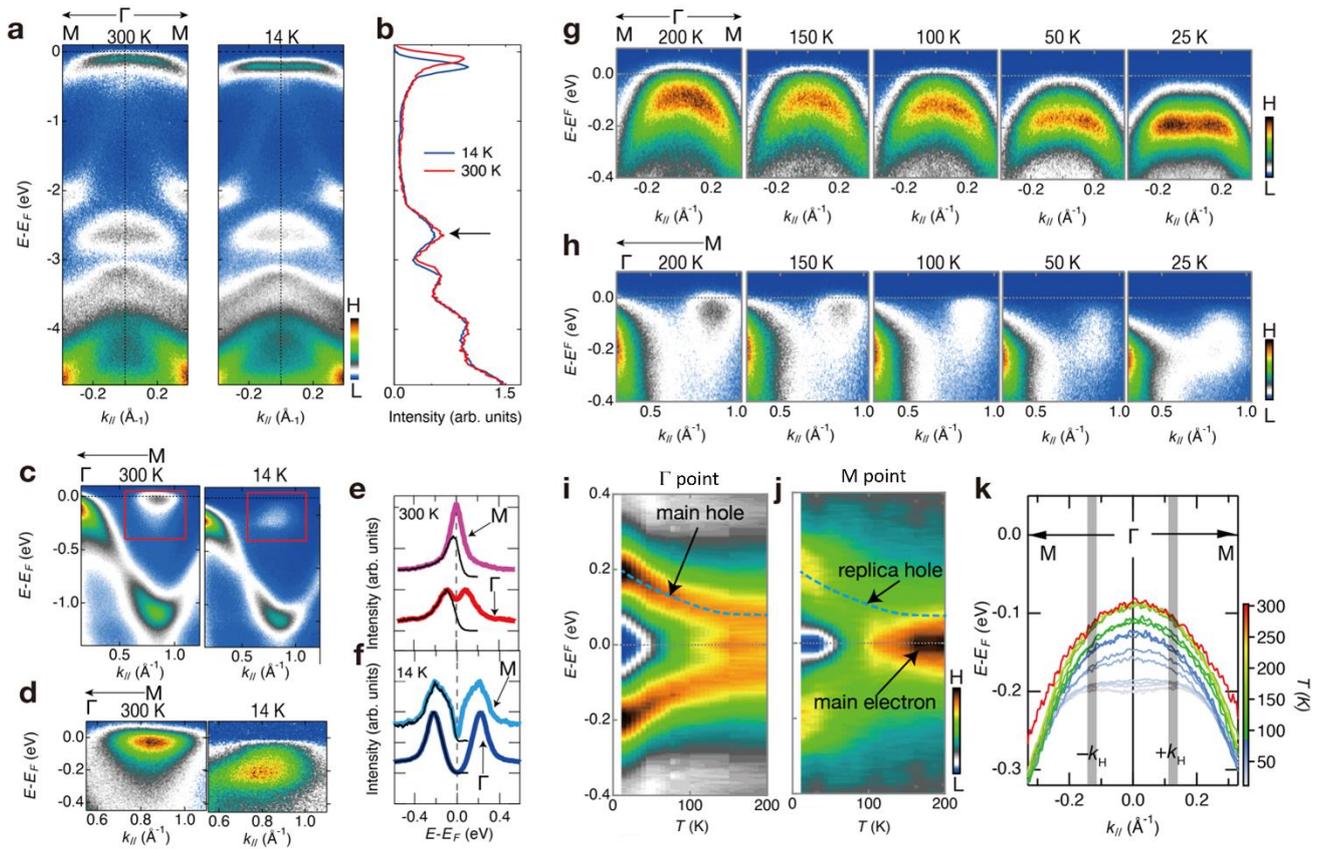

**Fig.2 | The ARPES measurement using a He-Iα ($h\nu$ = 20.218 eV) light source, focusing on temperature evolution from Γ to M point.**

**a,** The ARPES data for broad energy region near Γ point at 300 K (left) and 14 K (right). **b,** The energy distribution curve (EDC) at the Γ point for 300 (red) K and 14 K (blue). The shaded area represents significant band deformation, and the arrow indicates a slight change in spectral weight. See the main body for details. **c,** The low energy band structure near the M point at 300 K (left) and 14 K (right). **d,** The zoomed image of the band around the M point (see the highlighted area by the red line). The transition of band curvature from electron-like at 300 K (left) and hole-like at 14 K (right) is seen. **e, f,** The comparison of the symmetrized EDCs at the Γ and M points for 300 K **(e)** and 14 K **(f)**. **g, h,** The temperature evolution of the band around Γ **(g)** and M **(f)** points. **i, j,** The temperature evolution of the EDCs at Γ **(i)** and M **(j)**. The evolution from the main hole band dominates in **i**. The evolution from the shadow hole band and tail of the main electron band is seen in **j**. **i** and **j** are consistent with continuous connection between band at 300 K and 14 K shown in **a**, **c**, **d**. **k,** The temperature dependence of the dispersion for the main hole band along ΓM direction. The peak position is extracted from fitting energy and momentum/temperature dependent EDCs by the simple Lorentz function. The characteristic momentum $k_H$ (shaded momentum) highlights the emergence of a camelback-shaped band structure at low temperatures. See the main body for details.

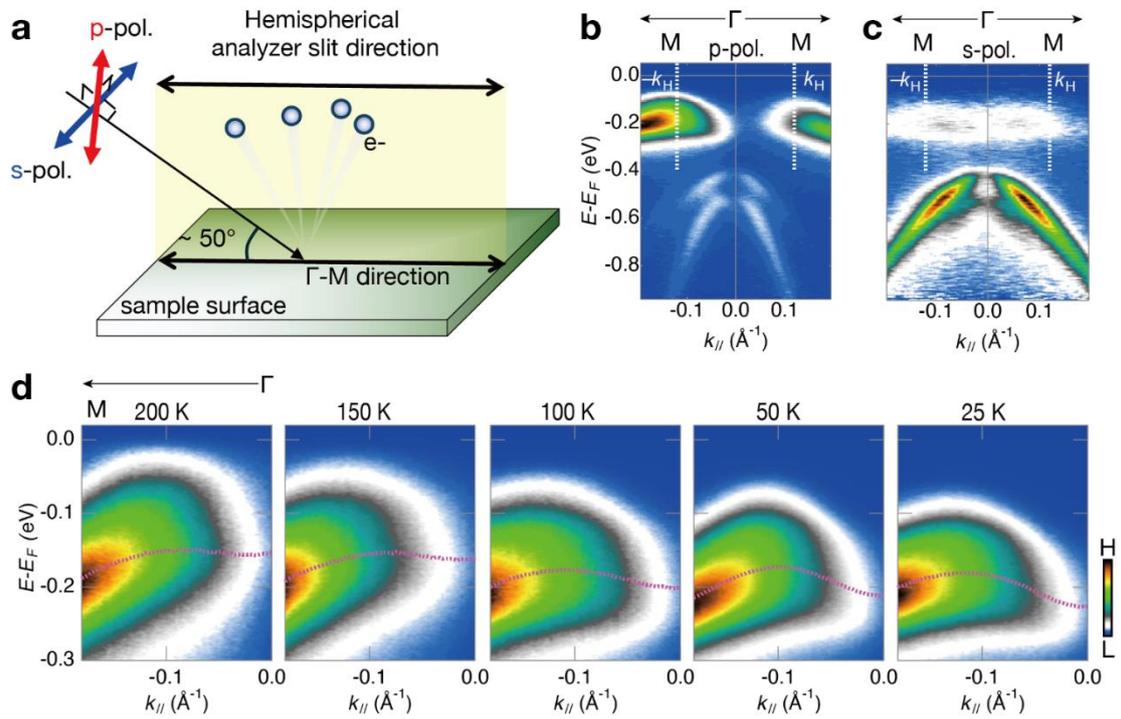

**Fig.3 | The ARPES measurement using a laser source (~7 eV), focusing on the camelback band around Γ the point.**
**a,** The set-up of the polarization data. The direction of s-polarization is perpendicular to the Γ–M ($a^*$) and out-of-plane direction ($c^*$), while the p-polarization direction is within the $a^*c^*$ plane and 50˚ tilted from $a^*$. **b, c,** The ARPES intensity along Γ-M at 7 K with p-polarization and s-polarization (**c**). **d,** The temperature evolution of the shape of the hole band, using p-polarization. The broken lines are a guide to the eye for dispersion. Positive momentum is not showing here due to intrinsic dramatic intensity reduction across the Γ point.

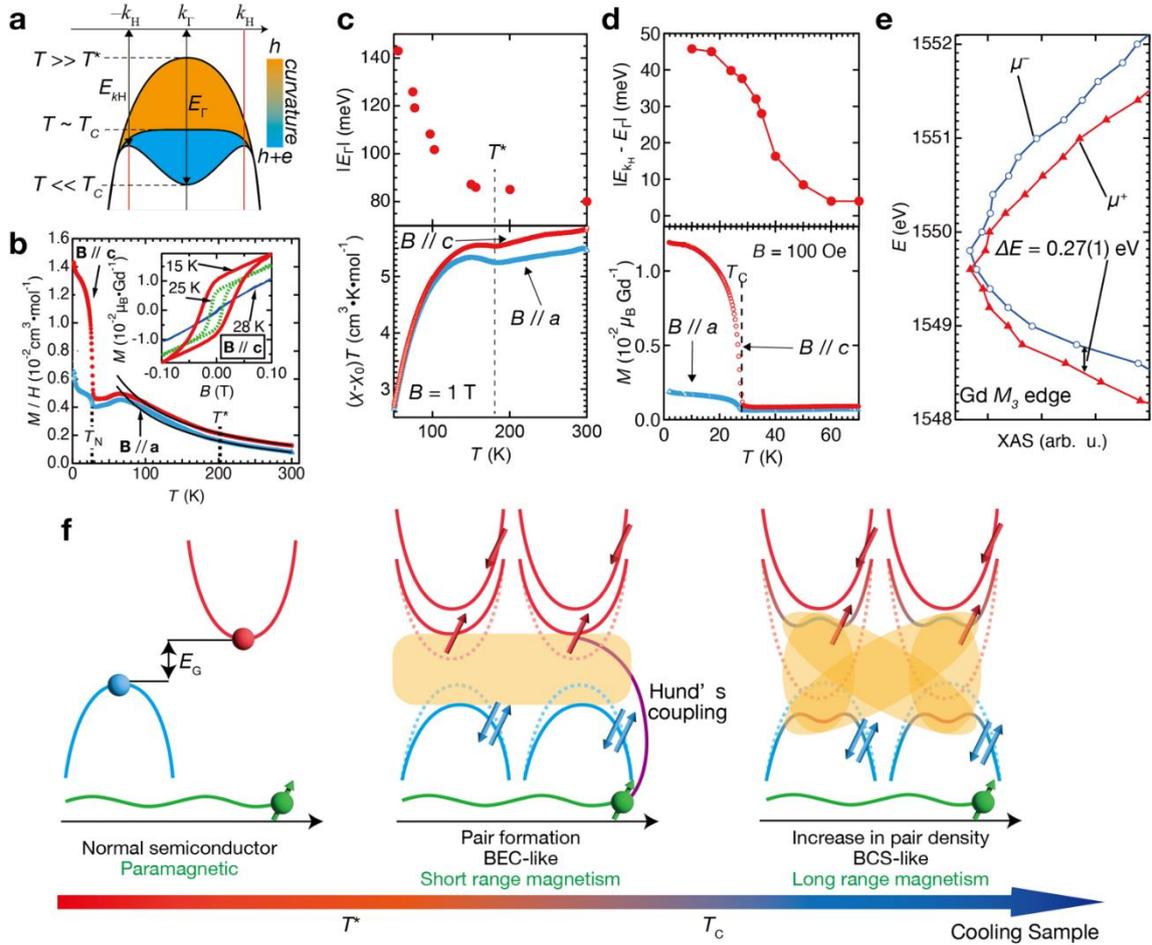

**Fig. 4 | The bridging between magnetic evolution (localized) and near EF excitonic band deformation (delocalized), through Hund's rule coupling.**

**a** The schematic for temperature evolution of band deformation near the Γ point. The schematic emphasizes two characteristic temperatures, $T^*$ and $T_c$. Consequently, the three characteristic temperature regions exist. For $T > T^*$, the hole band shape is independent of temperature. For $T_c < T < T^*$ (orange shaded area), the hole band top gradually shifts towards lower energy, whereas the signature of electron-like band curvature is not clearly seen in the hole band. For $T \ll T_c$ (blue shaded area), the camelback-shaped band shape dominates, which means electron-hole band hybridization is clearly reflected in k-space. See the main body for details. **b**, The temperature dependence of spin susceptibility in the *a* (in-plane) and *c* (out-of-plane) directions measured at $B = 1$ T with field cooling. The inset shows the magnetization process in the *c* direction below and above $T_c$. The black line indicates Curie-Weiss (CW) fit from a high-temperature region. The upward deviation is seen across $T^*$. **c**, The comparison between the temperature dependence of the band edge at Γ point $|E_\Gamma|$ (upper panel) and the reduced spin susceptibility $(\chi-\chi_0)T$ (lower panel). The high-temperature saturated value $\chi_0$ is extracted as a background in this figure. Note that $\chi_0$ is from the temperature region where CW fit captures experimental $\chi$ (**b**). **d**, The comparison between the temperature dependence of the degree of the camelback-shaped band formation $|E_\Gamma - E_{kH}|$ (upper panel) and the magnetization $M$ (lower panel). Here, the data in (c) is from the He discharge lamp, whereas (d) is from the *p*-polarized laser ARPES. **e**, The X-ray absorption spectroscopy (XAS) results at Gd $M_3$ edge with the photon helicity parallel ($\mu^+$, red) or antiparallel ($\mu^-$, blue) to the spin polarization at 15 K with the out-of-plane magnetic field of 1.9 T. From energy splitting between $\mu^+$ and $\mu^-$, the energy scale for spin splitting in delocalized Gd 5*d* band is estimated as $\Delta E = 270$ meV. **f**, The schematic of the excitonic band deformation in the BCS-BEC crossover region. There is an indirect band gap between 4*p* hole and 5*d* electron bands, both of which interact with each other through Coulomb interaction *U*. Because of Hund's rule coupling, 5*d* bands are spin-split with respect to the spin direction of 4*f* electrons. See the main body for details.

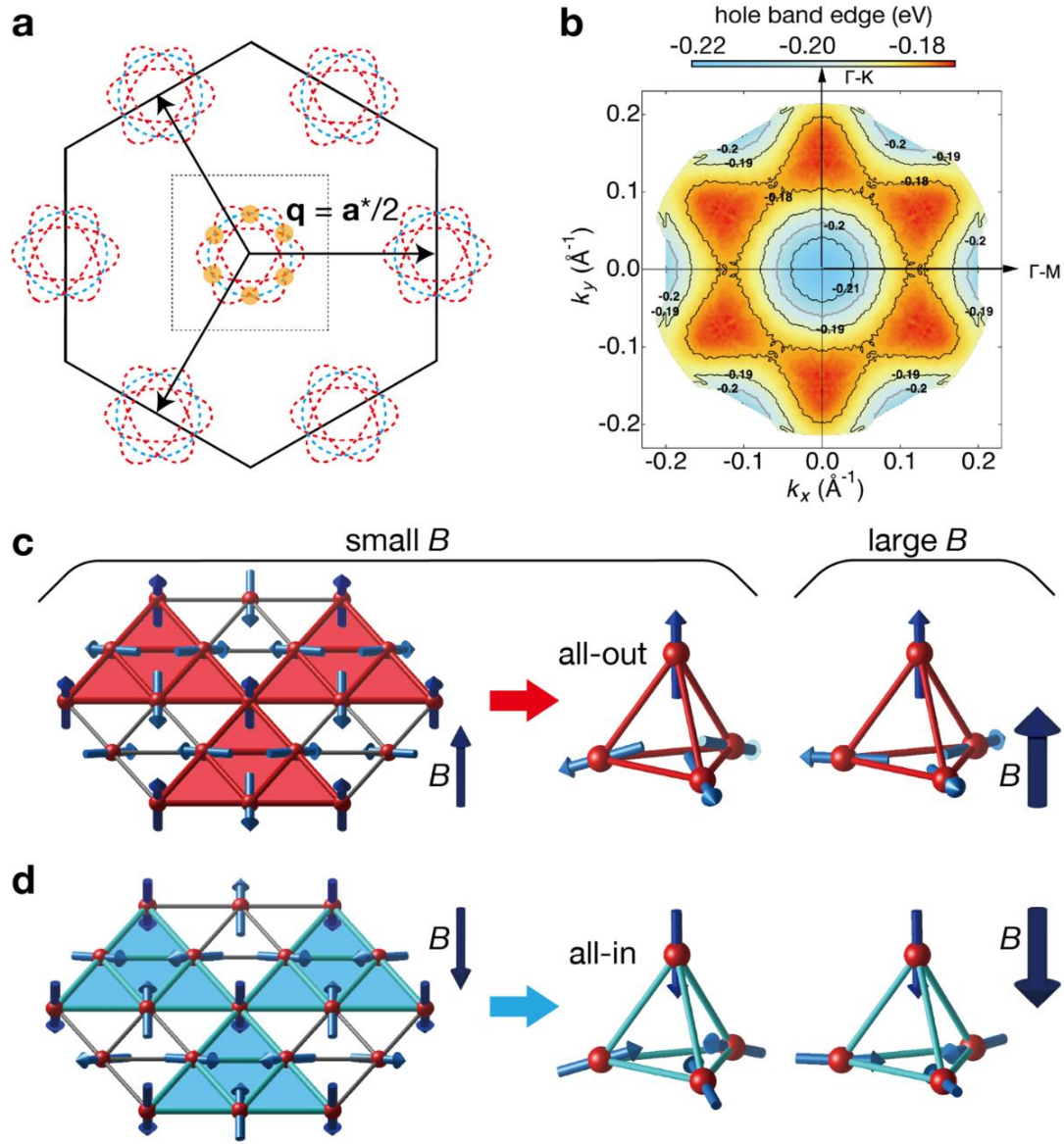

**Fig. 5 | Emergent topological magnetism from Hund's driven exciton condensation.**
**a**, The triple-$q$ exciton condensation leads to maximizing Fermionic energy gain. When the band top energy of unfolded hole constant energy contour is considered (blue), the degree of hybridization with triple-$q$ folded ellipse-shaped electron constant energy contour (red) becomes minimum along ΓM direction (shaded area around Γ). The region is highlighted as the orange-shaded area. **b**, The momentum dependence of the hole band top energy, from laser ARPES measurement at 7 K. In the measurement, *p*-polarization is used because of its intrinsic higher intensity but for $k < 0$ regions (see **Fig. 3a**). While the mapping is six-fold symmetrized from mapping for $k < 0$ areas, this mapping still supports triple-$q$ excitonic band hybridization (see **Supplementary Note 2**). **c, d,** The extracted triple-$q$ magnetic order (left), which represents a noncoplanar spin structure with ~2a length scale periodicity. Depending on the external field, spin structures can be fixed whether **c** or **d**. The spin arrangement in a closed loop of four sites (shown in red and blue bonds) can be mapped on a tetrahedron. The consequent all-out (top middle) and all-in (bottom middle) spin structures are topologically distinct objects. Larger magnetic fields tilt three spins on the bottom of each tetrahedron, which approaches nearly in-plane spin configurations as observed experimentally in the NMR measurement (right).

## Methods

### Crystal growth

Single crystals of GdGaI (GGI) were synthesized by high-temperature-solid-state reaction[15]. Handling of the elements and compounds is performed in an argon-filled glovebox. $GdI_3$ was made by the reaction of a stoichiometric amount of element. Gd and $I_2$ were sealed in an evacuated quartz tube in vacuo and reacted at 300°C for 24 hours followed by subsequent heating at 900°C for 24 hours to complete the reaction. The quartz tube was then placed vertically with one end outside the furnace and heated again at 900°C to facilitate the sublimation of polycrystalline $GdI_3$. White plate-like crystals of $GdI_3$ were formed at the colder end of the quartz tube. A stoichiometric amount of $GdI_3$, Gd, and Ga in a total weight of 1 g was sealed in a niobium tube with a diameter of 1 cm and length of 15 cm by an arc welding furnace under an argon atmosphere. The sealed Nb tube was placed in a quartz tube to prevent oxidation of Nb at elevated temperatures. The sealed quartz tube was placed horizontally in a three-zone furnace. The cold and hot end was kept 800°C and 1000°C, respectively. After a week, shiny golden platy crystals with a maximum dimension typically of 1 x 1 x 0.05 $mm^3$ formed at the hot end.

### Crystal characterization

Single crystal X-ray diffraction measurements were performed on a Rigaku SuperNova diffractometer at 298 K with Mo *Kα* source (λ = 0.71073 Å). A plate-like single crystal of GdGaI was fixed by paraffin oil on a rectangular glass capillary and the capillary was inserted in a round glass capillary in an argon-filled glovebox and sealed under vacuum. The X-ray data were collected and reduced using the CrysAlisPro software. Absorption corrections were made empirically. The initial model of the crystal structures was based on the earlier report based on powder diffraction data and refined using the program Olex2[25]. The result of the structural refinement is presented in **Extended Data Table 1**. For magnetometry measurement, using MPMS-3 (Quantum Design), a single crystal weighing 0.14 mg was placed on a diamagnetic sapphire substrate and coated with paraffin to protect sample degradation.

### Density Functional Theory Calculation

The first-principles calculations were performed using the Vienna Ab initio Simulation Package (VASP) [26,27] with the projector augmented wave (PAW)[28] potentials within the framework of density functional theory (DFT)[29]. The exchange-correlation function was treated within the Perdew-Burke-Ernzerhof (PBE) under generalized gradient approximations (GGA)[30]. Cutoff energy was set to 500 eV and structural relaxation was performed until the residual forces were no greater than $10^{-3}$ eV/Å, and the energy convergence criteria for self-consistency were set at $10^{-6}$ eV. A Γ-centered Monkhorst-Pack grid of 12 x 12 x 3 in the first Brillouin zone was used. Moreover, spin-orbit coupling (SOC) was considered in the band structure calculations, while the Gd-*f* orbital was considered only for the magnetic state. The optimized lattice constant is *a* = 4.229 Å and *c* = 12.386 Å for the non-magnetic state, while *a* = 4.247 Å and *c* = 12.439 Å for the ferromagnetic state. In this study, the main purpose of DFT calculation is to catch the minimal essences in experimental observations. Thus, implementing correlation effects, band folding, and detailed parameter tuning are left for future study. See **Supplementary Note 1** for further details.

### Angle resolved photoemission spectroscopies

In this report, we employed three types of ARPES experiments with different light sources (He discharge lamp, laser, and synchrotron soft X-ray). In all ARPES measurements, the sample surface was obtained by cleaving in ultra-high vacuum (UHV). Intrinsic cleavable nature allows facile access to high-quality surfaces. The position of SX-ARPES is to confirm the bulk electronic nature and degree of $k_z$ dispersion, while energy resolution is relatively low compared to other ARPES studies. The SX-ARPES spectra were collected by the DA30 photoelectron analyzer (Scienta Omicron) with setting a total energy resolution to 50 ~100 meV. The length scale of spot size ~ 10 x10 $μm^2$. The position of He-based ARPES is to confirm electronic structure consistently between Γ and M points, with energy resolution better than SX-ARPES. The instrumental resolution is < 10 meV, and the spot size is ~ 0.5 mm. Focusing on near Γ point, the position of laser-ARPES is to unveil electronic state as high momentum/energy resolution as possible. Also, polarization dependence provides useful information to understand momentum evolution of orbital characters. We use p-polarization and s-polarization of incident 7 eV laser[31]. See **Supplementary Note 2** for further details. The data in **Fig. 4c** is from He discharge lamp (shown in **Fig. 2**), whereas **Fig. 4d** is from p-polarization laser ARPES (shown in **Fig. 3**). The degree of camelback band formation between the He discharge lamp (**Fig. 2k**) and laser (**Fig. 3d**) based ARPES measurements show a slight difference, presumably from distinct energy/momentum resolution and finite $k_z$ dispersion (potential temperature dependent). However, the characteristic temperatures around $T_c$ and $T^*$ extracted from both light sources are similar, and therefore, we would like to position this mismatch as a minor factor within the main purpose of this study.

**X-ray magnetic circular dichroism**
The X-ray absorption spectrum (XAS) and X-ray magnetic circular dichroism measurement (XMCD) measurements were performed at the helical undulator beamline BL25SU of SPring-8[32]. The monochromator resolution $E/\Delta E$ was over 5000 at Gd $M_3$-edge (~1.55 keV). Experimental setup is shown in **Extended Data Fig. 1a**. The designed beam spot size was 100-200(H) × 10(V) μm$^2$. For the XMCD measurements, absorption spectra for circularly polarized X-rays with the photon helicity parallel ($\mu^+$) or antiparallel ($\mu^-$) to the spin polarization were recorded in the total-electron-yield (TEY) mode obtained by the sample drain current. The $\mu^+$ and $\mu^-$ spectra were taken under both positive and negative applied magnetic fields and averaged to eliminate experimental errors. All XAS and XMCD spectroscopy measurements were performed at 1.9 T, the maximum applied magnetic field in the system[33]. To measure the perpendicular magnetic moment, the angles of the external magnetic fields measured from the sample surfaces were set to 90°. The incident X-rays were tilted at 10° to the magnetic field. The in-situ cleavage obtained the clean surface at room temperature and ~10$^{-6}$ Pa. The measurement temperature was 15 K. **Extended Data Figs. 1b-h** shows the result of XAS and XMCD. XMCD spectrum of $M_{4,5}$ has a complex shape due to $4f^7$ multiplets while the shape of $M_{4,5}$ is characteristic of $5d^0$ configuration. XMCD intensity of $M_5$ edge is two orders of magnitude larger than that of $M_3$ edge, suggesting bulk magnetization is dominated by $4f$ spins. Fitting of $\mu^+$ and $\mu^-$ resulted in estimated band splitting of 0.27(1) eV as shown in **Extended Data Fig. 2**. As the magnetic dipole term $T_z$ contributes to the energy shift in addition to spin magnetic moment term, the estimated energy splitting corresponds to a lower bound.

**Nuclear Magnetic Resonance**
$^{127}$I-NMR measurement (**Extended Data Fig. 3a**) was performed for four pieces of single crystals stacked along the $c$ axis and covered with paraffin to prevent contamination by air. The crystals were aligned sufficiently well to the c axis supported by the flat and plate-like sample shape. While in-plane orientation was not aligned, this does not impact the main conclusion of this study. The external magnetic field was applied along the $c$ direction, which is perpendicular to the flat sample surface. The field-sweep $^{127}$I-NMR spectrum was measured by recording the NMR intensity at a fixed resonant frequency of 95.51 MHz during the field sweep. The reference field is determined by the gyromagnetic ratio of $^{127}$I nuclear moment, $g_l$ = 8.557 MHz/T, and is 11.162 T as indicated by the vertical dashed line in **Extended Data Fig. 3b**. The $^{127}$I-NMR spectrum was obtained with a reasonably good signal-to-noise ratio at the base temperature of 1.4 K. The signal intensity decreases at higher temperatures following the decrease in the nuclear magnetization by the Boltzmann factor, and thus we cannot measure the NMR spectrum in the paramagnetic state above long-range magnetic transition temperature.

**Symmetry analysis**
Symmetry analysis of the magnetic structure was performed by ISODISTORT software[34,35]. In GdGaI at room temperature, Gd occupies $2c$ site of the space group $P$-$3m$1. ARPES measurement indicates that the likely magnetic propagation vector is $k$ = (1/2, 0, 0). There are three arms of the star of $k$ = (1/2, 0, 0), comprising single-$q$, double-$q$ and triple-$q$ structures. Among twelve symmetry operations in $P$-$3m$1, namely {$I$, $-I$, $m_{100}$, $2_{100}$, $m_{010}$, $2_{010}$, $m_{110}$, $2_{110}$, $3^+_{001}$, $-3^+_{001}$, $3^-_{001}$, $-3^-_{001}$}, the following symmetry operations keep the set of wavevectors invariant: {$I$, $-I$, $m_{010}$, $2_{010}$} for $k$ = (1/2, 0, 0), {$I$, $-I$, $m_{110}$, $2_{110}$} for $k$ = (1/2, 0, 0) and (1/2, 1/2, 0) and all the symmetry operations for $k$ = (1/2, 0, 0), (1/2, 1/2, 0), and (0, 1/2, 0). Here the $I$, $-I$, $m_{hkl}$, $2_{hkl}$, $3^±_{hkl}$ and $-3^±_{hkl}$ correspond to identity, space inversion, mirror perpendicular to a vector $r$ = $h$**a** + $k$**b** + $l$**c**, two-fold rotation about $r$, (counter) clockwise three-fold rotation about $r$, and (counter) clockwise three-fold roto-inversion about $r$, respectively. List of symmetry-allowed single-$q$, double-$q$, and triple-$q$ magnetic structures are provided in **Extended Data Table 2**. Among the candidate magnetic structures, only All-out$^+$ structures allow net out-of-plane moment, which has the point group symmetry of -$3m$'1 (**Extended Data Figs. 3c-d**). Group-subgroup relations of all the magnetic space groups and the schematic images are shown in **Extended Data Fig. 4**. As explained later, the effect of out-of-plane magnetic field is weak perturbation to the spin arrangement causing gradual canting of spins because of no field induced phase transitions below the measurement field of 11 T in NMR experiments. **Extended Data Table 3** lists the internal field at the iodine site exerted from the three nearest Gd atoms in the presence of out-of-plane magnetic field, which is assumed to deviate from the magnetic structure from the **Extended Data Table 2**. The observation of two types of internal fields with unequal population of 1 : 2.4 is consistent with the All-out+ structure, which yields the distribution of 2 : 6 = 1 : 3, but cannot be explained by any single-$q$ structures because they show either single site or two sites with equal population. Double-$q$ structures generally have three types of distinct internal fields. Regarding the possibility of orders with other $q$ vectors, the observed asymmetric peak distribution eliminates the possibility of orders in all the other special positions in the Brillouin zone. Commensurate spiral structures at general positions such as (1/$l$, 1/$m$, 1/$n$) with $l$, $m$, and $n$ being integer are also excluded because they would have more than two types of the internal fields. As the current NMR measurement detects only the out-of-plane component, it is restricted to the All-out+ type while the in-plane ones can in principle acquire additional phase factor in the $k_z$ direction keeping the noncoplanar tetrahedral geometry in each layer. Such structures are unlikely because of the cost of interplane coupling compared to the All-out+ structure. It is noted that a similar triple-$q$ magnetic structure is observed in $Co(Nb,Ta)_3S_6$, which has a large fermi surface and exhibits substantial anomalous Hall effect [36,37].

**High field magnetization measurement**

Magnetization at pulsed magnetic fields up to 46 T was measured at 1.4 K and 4.2 K using polycrystalline sample at MegaGauss Science Laboratory in the University of Tokyo (**Extended Data Fig. 5**). The absolute value of magnetization was calibrated by the data obtained from MPMS-3 using the data at 4.2 K. Magnetization saturates at 7 μ$_B$ as expected for $Gd^{3+}$ ($S$ = 7/2). Assuming only canting of spins on the Gd$_2$ site in the all-out+ structure, the tilting angle is estimated as –2° away from the $ab$ plane towards the $c$-axis, which amounts to magnetization of 1.7 μ$_B$/Gd and consistent with the powder averaged value of 1.6 μ$_B$ at 11.162 T (see broken lines in **Extended Data Fig. 5**). No field-induced transition was observed below 15 T.

| Chemical formula | GdGaI |
|---|---|
| Space group | $P$-$3m1$ |
| $a$ (Å) | 4.1985(3) |
| $c$ (Å) | 11.476(1) |
| $V$ (Å$^3$) | 175.19(2) |
| $Z$ | 2 |
| $\mu$ (mm$^{-1}$) | 34.968 |
| $R_{p,observed}$ | 3.90 |
| $R_{p,all}$ | 5.48 |
| $R_{wp,observed}$ | 7.24 |
| $R_{wp,all}$ | 8.33 |
| $R_{int}$ | 5.18 |
| Index range | -4 < $h$ < 4 |
| | -5 < $k$ < 5 |
| | -12 < $l$ < 14 |
| $N_{measured}$ | 1199 |
| $N_{unique}$ | 214 |
| $N_{observed}$ | 177 |
| $N_{param}$ | 11 |
| Crystal size ($\mu$m$^3$) | 50×50×10 |
| Extinction coefficient | 0.0104(17) |

| Atom | Wyckoff position | $x$ | $y$ | $z$ | $U_{iso}$ | $U_{11}$ | $U_{22}$ | $U_{33}$ | $U_{12}$ | $U_{13}$ | $U_{23}$ |
|---|---|---|---|---|---|---|---|---|---|---|---|
| Gd | 2$c$ | 0 | 0 | 0.1781(1) | 9.4(5) | 7.5(6) | 7.5(6) | 13.0(1) | 3.8(3) | – | – |
| Ga | 2$d$ | 2/3 | 1/3 | 0.0273(3) | 9.6(8) | 4.6(9) | 4.6(9) | 20(2) | 2.3(5) | – | – |
| I | 2$d$ | 1/3 | 2/3 | 0.3528(2) | 12.7(6) | 11.4(7) | 11.4(7) | 15(1) | 5.7(4) | – | – |

**Extended Data Table 1 | Crystallographic data and the result of structural refinement for GdGaI from single-crystal X-ray data at 298 K.**
Observed peaks satisfy $I > 2\sigma(I)$. Atomic fractional coordinates, and equivalent isotropic $U_{iso}$ and anisotropic $U_{ij}$ displacement parameters (in units of 10$^{-3}$Å$^2$) with estimated standard deviations in parentheses. Equivalent isotropic displacement parameter is defined as one third of the trace of $U_{ij}$.

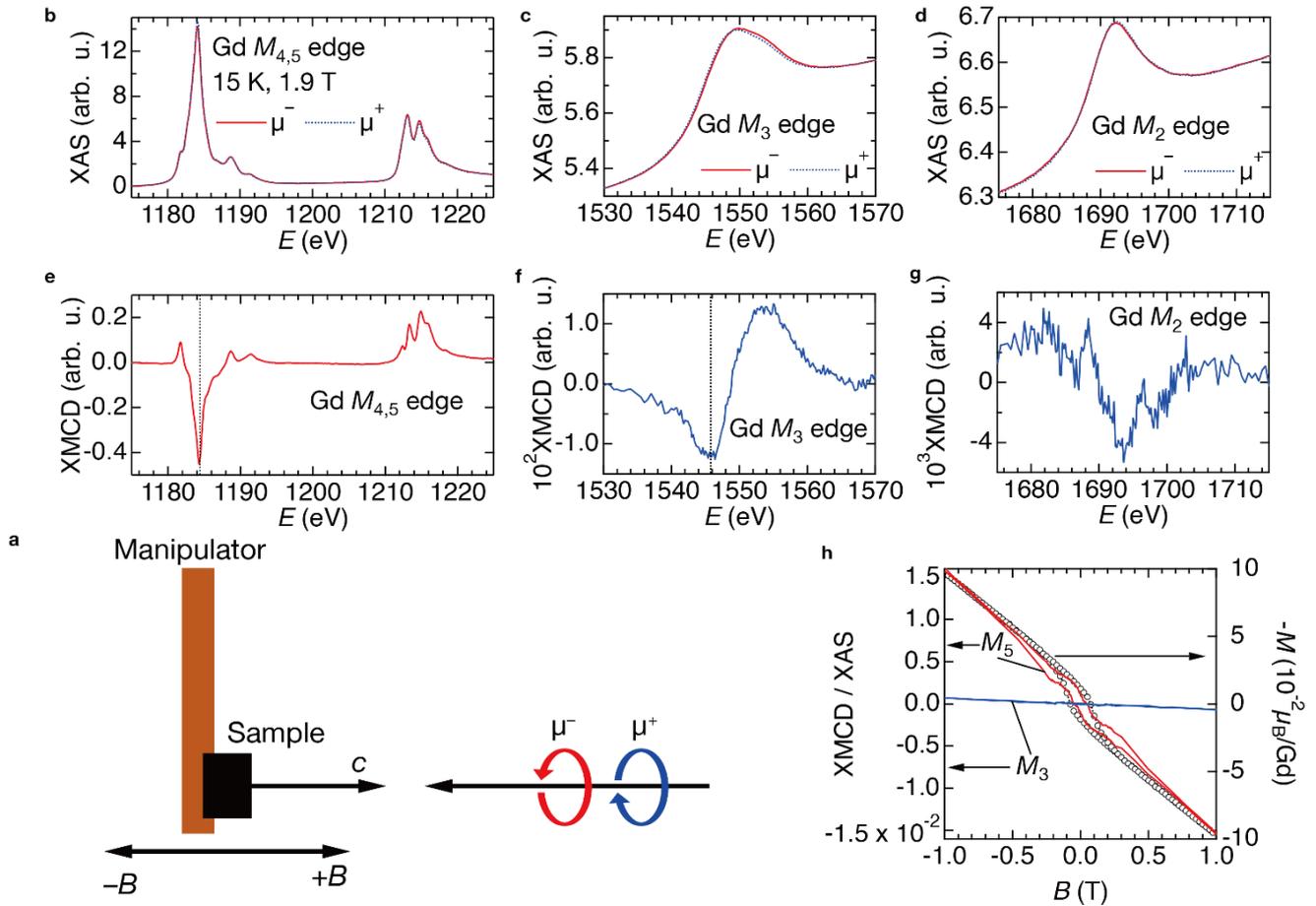

**Extended Data Fig. 1 | XAS and XMCD experiment.**
**a**, Measurement setup. **b-d**, XAS at $M_{4,5}$ edges ($3d_{3/2,5/2} \rightarrow 4f$) and $M_{2,3}$ edges ($3p_{1/2,3/2} \rightarrow 5d$) with the photon helicity parallel ($\mu^+$) or antiparallel ($\mu^-$) to the spin polarization. **e-g**, XMCD at $M_{4,5}$ and $M_{2,3}$ edges. The dashed lines correspond to the energy used for the field sweep measurement of the XMCD intensity shown in the panel **h**. **h**, Comparison of XMCD intensity measured at 15 K with the magnetization along the $c$-axis measured at 10 K. In Fig. 4e, the $\mu^-$ curve is multiplied by 0.99924 to clearly show the energy shift (see the Extended Data Fig. 2)

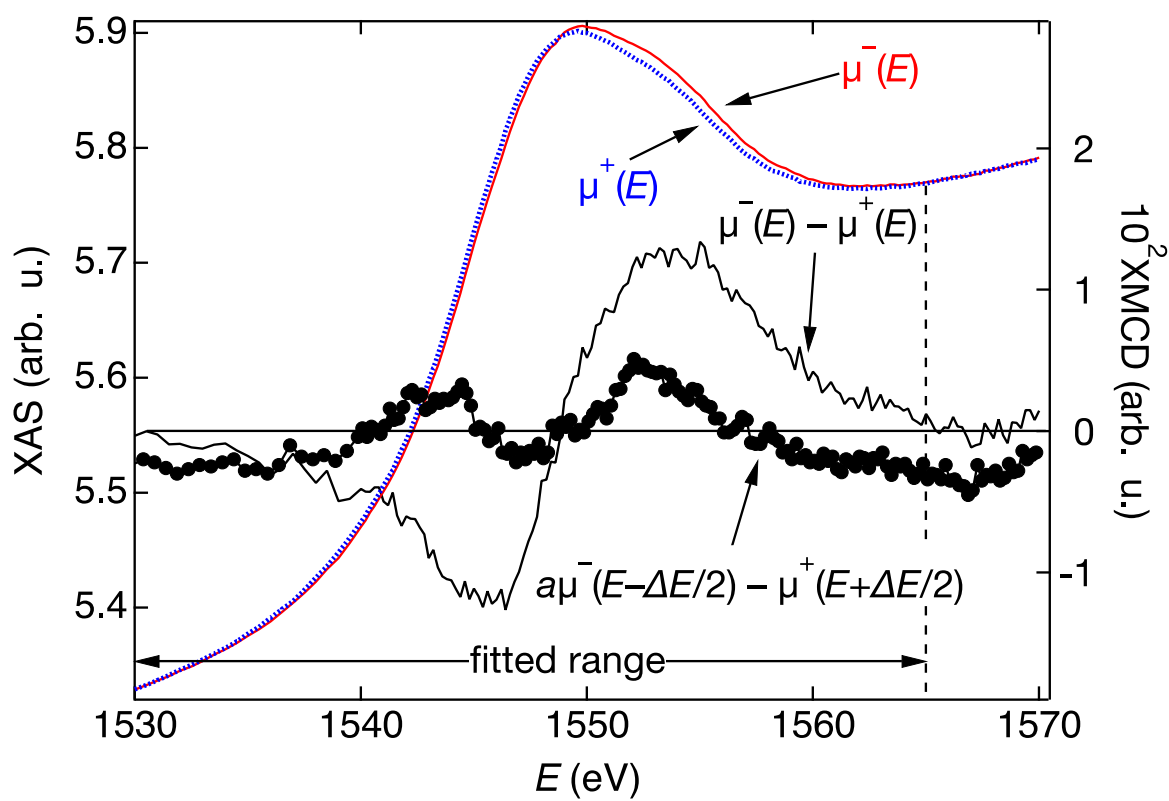

**Extended Data Fig. 2 | Splitting of 5d bands estimated at the $M_3$ edge.**
The 5d spin-up band is shifted to lower energy compared to the spin-down band and partial occupation of the spin up band at the fermi level leads to the value of $\mu^-$ larger than that of $\mu^+$. To estimate the spin splitting, $\mu^+(E+\Delta E/2)$ is fitted to $a\mu^-(E-\Delta E/2)$, where $a$ is the scale factor slightly smaller than 1 and $\Delta E$ is the energy splitting between spin up and down bands. The fitting converged to $a = 0.99924(4)$ and $\Delta E = 0.266(7)$ eV when the fitting range is between 1530 and 1565 eV; the residual is plotted as black filled circles with a line. By changing the fitting range, the energy shift including the uncertainty due to the fitting range is estimated as 0.27(1) eV.

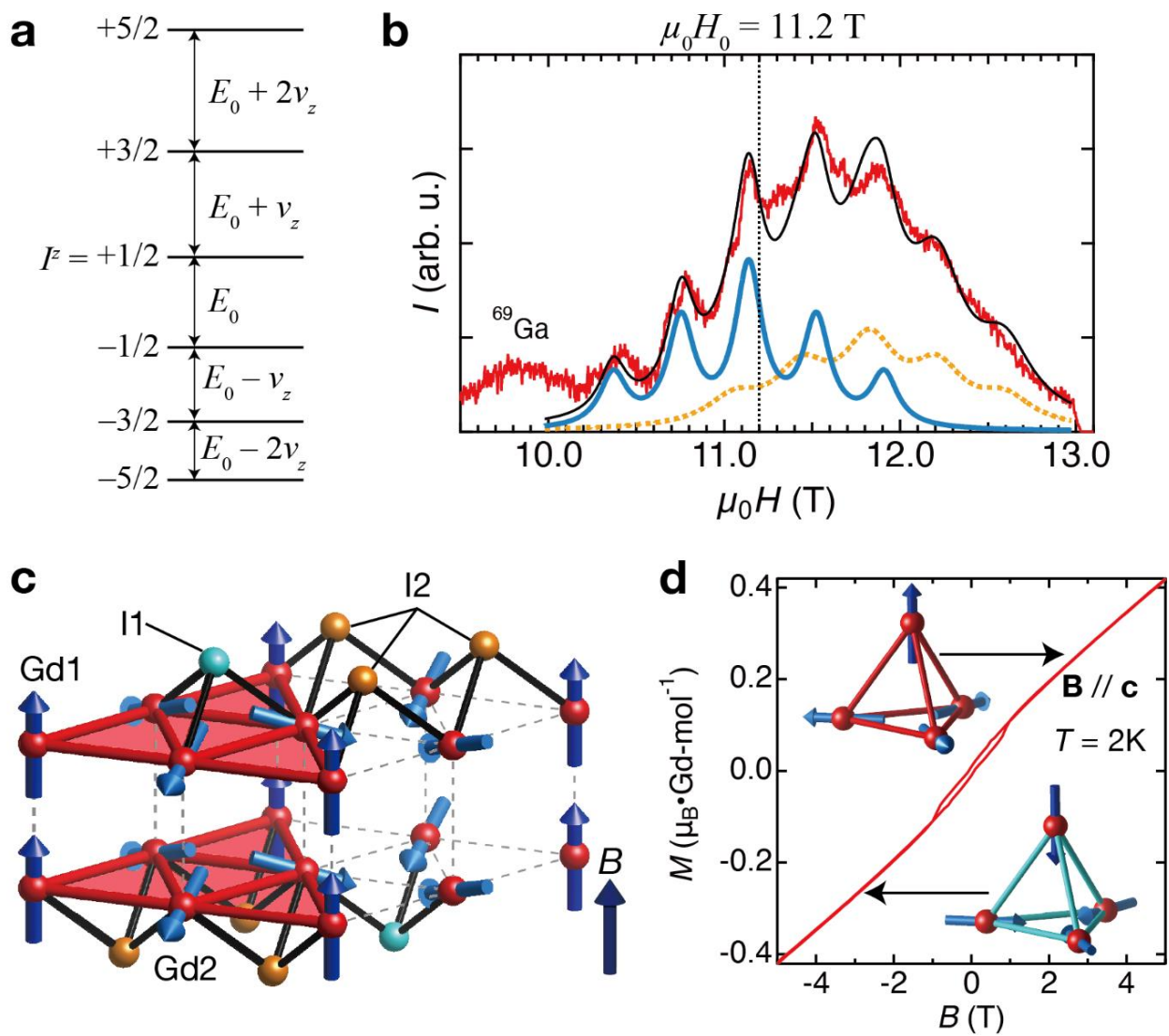

**Extended Data Fig. 3 | NMR measurements on GGI.**

**a,** Nuclear spin transition responsible for a NMR spectrum of $^{127}$I. In the presence of three-fold symmetry, Zeeman term $E_0$, which is proportional to the sum of applied field and internal field due to magnetic ions, and quadrupole interaction $v_z$ determine the energy separation between the six states specified by angular momenta. As the NMR transition happens between two states with relative difference in $I^z$ by 1, five peaks are observed. **b,** NMR spectrum of GdGaI at 1.4 K. The red and black lines indicate the observed and simulated spectrum. The sky-blue thick and orange dotted lines are contribution of two magnetically inequivalent iodine sites to yield the black curve. The broad hump below 10 T is attributed to NMR spectrum of $^{69}$Ga. **c,** Solution of the magnetic structure from the NMR measurement. The noncoplanar magnetic structure is described by three propagation vectors, $q$ = (1/2, 0, 0) and (0, -1/2, 0), and (-1/2, 1/2, 0). Red spheres and blue arrows represent Gd atoms and magnetic moments, respectively. The iodine sites with different magnetic symmetries are shown in different colors, namely orange and sky blue. There are two types of iodine atoms in the magnetic unit cell, of which number is two and six for I1 and I2 site and internal field is negative and positive, respectively. **d,** Evolution of the magnetic structure under $c$-axis magnetic fields at 2 K in comparison with the magnetization process. In the All-out[+] structure, three spins on the Gd2 site are tilted to the field direction to approach the in-plane spin arrangement whereas the spins on Gd1 site are fixed in the c direction, leading to suppression and enhancement of the internal fields at iodine sites I1 and I2, respectively. The hysteresis near zero field indicates the coexistence of the all-out[+] structures with opposite chirality, which are switched by the sign of magnetic fields.

| q | Notation | M. S. G. | (a', b', c') | Origin | Atom | W. P. | (x, y, z) | M |
|---|---|---|---|---|---|---|---|---|
| $q_1$ | Stripe-**x**$^+$ | 10.49 $P_C2/m$ | $2a+b, b, c$ | (0, 0, 0) | Gd1 | 4$i$ | (0, 0, $z_{Gd}$) | $S(-1/2, \sqrt{3}/2, 0)$ |
| | Stripe-**yz**$^+$ | 14.83 $P_A2_1/c$ | $c, b, -2a-b$ | (0, 0, 0) | Gd1 | 4$i$ | ($z_{Gd}$, 0, 0) | $S(\sqrt{3}/2\sin\theta, 1/2\sin\theta, \cos\theta)$ |
| | Stripe-**yz**$^-$ | 13.73 $P_A2/c$ | $c, b, -2a-b$ | (-1/2, 0, 0) | Gd1 | 4$i$ | ($z_{Gd}$, 1/4, 1/4) | $S(\sqrt{3}/2\sin\theta, 1/2\sin\theta, \cos\theta)$ |
| | Stripe-**x**$^-$ | 11.57 $P_C2_1/m$ | $2a+b, b, c$ | (1/2, 1/2, 0) | Gd1 | 4$i$ | (3/4, 3/4, $z_{Gd}$) | $S(-1/2, \sqrt{3}/2, 0)$ |
| $q_1$ $q_2$ | Double-$q^+$ | 12.64 $C_a2/m$ | $-2a-4b, 2a, c$ | (0, -1, 0) | Gd1 | 8$m$ | (1/2, 0, $z_{Gd}$) | $S(0, 1, 0)$ |
| | | | | | Gd2 | 8$n$ | (1/4, 1/4, $z_{Gd}$) | $S'(0, \sin\theta, \cos\theta)$ |
| | Double-$q^-$ | 12.64 $C_a2/m$ | $-2a-4b, 2a, c$ | (3/2, 1, 0) | Gd1 | 8$m$ | (1/4, 0, $z_{Gd}$) | $S(0, 1, 0)$ |
| | | | | | Gd2 | 8$n$ | (1/2, 1/4, $z_{Gd}$) | $S'(\sin\theta, 0, \cos\theta)$ |
| $q_1$ $q_2$ $q_3$ | 120°$^+$ | 164.85 $P$-3$m$1 | $2a, 2b, c$ | (0, 0, 0) | Gd1 | 2$c$ | (0, 0, $z_{Gd}$) | (0, 0, 0) |
| | | | | | Gd2 | 6$i$ | (1/2, 1/2, $z_{Gd}$) | $S(1/2, \sqrt{3}/2, 0)$ |
| | All-out$^+$ | 164.89 $P$-3$m'$1 | $2a, 2b, c$ | (0, 0, 0) | Gd1 | 2$c$ | (0, 0, $z_{Gd}$) | $S(0, 0, 1)$ |
| | | | | | Gd2 | 6$i$ | (1/2, 1/2, $z_{Gd}$) | $S'(\sqrt{3}/2\sin\theta, -1/2\sin\theta, -\cos\theta)$ |
| | All-out$^-$ | 164.88 $P$-3$'m'$1 | $2a, 2b, c$ | (0, 0, 0) | Gd1 | 2$c$ | (0, 0, $z_{Gd}$) | $S(0, 0, 1)$ |
| | | | | | Gd2 | 6$i$ | (1/2, 1/2, $z_{Gd}$) | $S'(\sqrt{3}/2\sin\theta, -1/2\sin\theta, -\cos\theta)$ |
| | 120°$^-$ | 164.87 $P$-3$'m$1 | $2a, 2b, c$ | (0, 0, 0) | Gd1 | 2$c$ | (0, 0, $z_{Gd}$) | (0, 0, 0) |
| | | | | | Gd2 | 6$i$ | (1/2, 1/2, $z_{Gd}$) | $S(1/2, \sqrt{3}/2, 0)$ |

**Extended Data Table 2 | List of the magnetic structures described by propagation vectors of $q_1$ = (1/2, 0, 0), $q_2$ = (0, 1/2, 0) and $q_3$ = (1/2, 1/2, 0) in the maximal subgroup of $P$-3$m$11'.**
M. S. G. and W. P. and $S$ indicate the magnetic space group, Wyckoff position, and magnetic moment. The number in the Wyckoff positions represents that of iodine sites in the unit cell. The magnetic moment at Gd site $M$ is defined with respect to the Cartesian coordinate of the original unit cell [**a**/$a$, (2**a**+**b**)/($\sqrt{3}a$), **c**/$c$]. Notation of each magnetic structure with either positive or negative suffix indicates that the magnetic moment at ($x$, $y$, $z$) and ($x$, $y$, –$z$) is identical or antiparallel, respectively. The moment size of each Gd atom is assumed to be $S$ and $S'$, which are expected to be close to 7$\mu_B$, except for one of the site in the "120°$^\pm$" structures. Here $\theta$ can take arbitrary real values. A net out-of-plane net moment appears only in the case of All-out$^+$ structure with $\cos\theta \neq -1/3$.

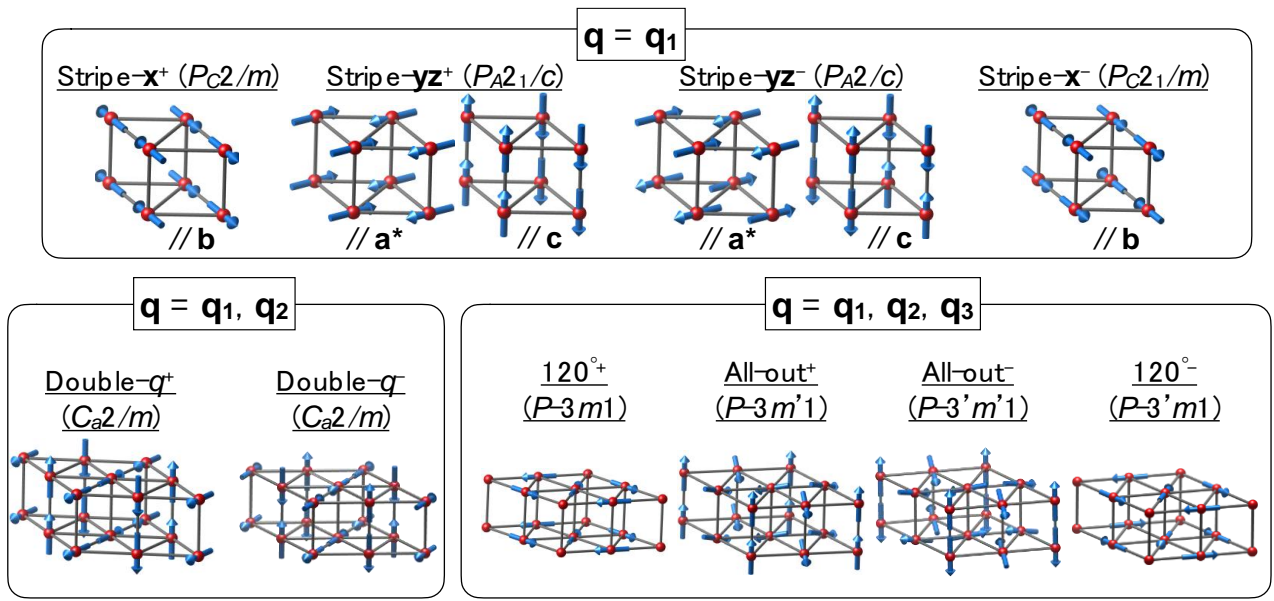

**Extended Data Fig. 4 | Graphical image of the magnetic structures described in Extended Data Table 2.**

| Label | $B_{loc}$ | $N_I$ |
|---|---|---|
| Stripe-**x**$^+$ | $\delta_1$ | 4 |
| Stripe-**yz**$^+$ | $S\cos\theta+\delta_1$ | 2 |
|  | $-S\cos\theta+\delta_2$ | 2 |
| Stripe-**yz**$^-$ | $S\cos\theta+\delta_1$ | 2 |
|  | $-S\cos\theta+\delta_2$ | 2 |
| Stripe-**x**$^-$ | $\delta_2$ | 4 |
| Double-$q^+$ | $S\cos\theta+\delta_1$ | 4 |
|  | $\delta_2$ | 8 |
|  | $-S\cos\theta+\delta_3$ | 4 |
| Double-$q^-$ | $S\cos\theta+\delta_1$ | 4 |
|  | $\delta_2$ | 8 |
|  | $-S\cos\theta+\delta_3$ | 4 |
| All-out$^+$ | $-3S\cos(\theta-\delta_1)$ | 2 |
|  | $S(1-2\cos(\theta-\delta_1))$ | 6 |
| All-out$^-$ | $-3S\cos(\theta-\delta_1)$ | 1 |
|  | $S(-1+2\cos(\theta-\delta_1))$ | 3 |
|  | $S(1-2\cos(\theta+\delta_2))$ | 3 |
|  | $+3S\cos(\theta+\delta_2)$ | 1 |

**Extended Data Table 3 | Internal field $B_{loc}$ and the number of iodine atoms in the magnetic unit cell $N_I$ under external out-of-plane magnetic field.**

The number in the Wyckoff positions represents that of iodine sites in the unit cell. The magnetic field allows for canting of each Gd magnetic moment. "120°±" structures listed in **Extended Data Table 2** are excluded here because of the presence of null moment size at one of the sites ($2c$ position in the zero field). Here $\delta_i$ ($i$ = 1, 2, 3) are small real values corresponding to spin canting.

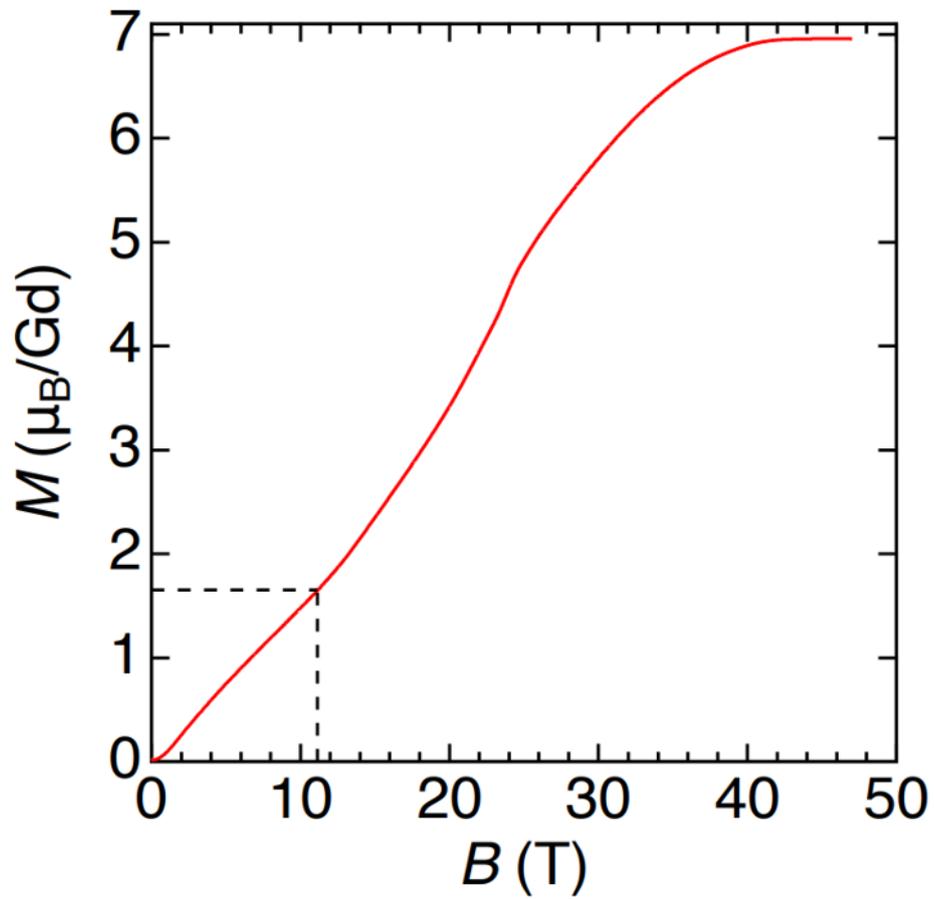

**Extended Data Fig. 5 | High field magnetization measurements on polycrystalline GdGaI at 1.4 K.**
The red line represents the up-sweep magnetization process. The magnetization at 11.162 T is shown by the dotted line.